\documentclass[5p]{elsarticle}

\usepackage{amsmath}
\usepackage{natbib}

\journal{Acta Materialia}

\usepackage{amsmath,amssymb}
\usepackage{graphicx}
\usepackage{bm}
\usepackage{epstopdf}
\usepackage{color}

\newcommand{\co}        {$^{59}$Co}
\newcommand{\He}		{He$^{+}$}
\newcommand{\cms}		{Co$_2$MnSi}

\begin{document}

\begin{frontmatter}

\title{Structure-property relationship of Co$_2$MnSi thin films in response to \He -irradiation }

\author[IFW]{F. Hammerath}
 \ead{f.hammerath@ifw-dresden.de}
\author[HZDR]{R. Bali}
\author[HZDR]{R. H\"ubner}
\author[IFW]{M. R. D. Brandt}
\author[IFW]{S. Rodan}
\author[HZDR]{K. Potzger}
\author[HZDR]{ R. B\"ottger}
\author[NIMS]{Y. Sakuraba}
\author[IFW,TU]{B. B\"uchner}
\author[IFW,TU]{S. Wurmehl}

\address[IFW]{IFW-Dresden, Institute for Solid State Research, Helmholtzstra{\ss}e 20, 01069 Dresden, Germany}
\address[HZDR]{Helmholtz-Zentrum Dresden-Rossendorf, Institute of Ion Beam Physics and Materials Research, Bautzner Landstrasse 400, 01328 Dresden, Germany}
\address[NIMS]{National Institute of Materials Science, Tsukuba, Ibaraki, Japan}
\address[TU]{Institute of Solid State and Materials Physics, TU Dresden, 01062 Dresden, Germany}

\begin{abstract}
We investigated the structure-property relationship of \cms\ Heusler thin films upon the irradiation with \He\ ions. 
The variation of the crystal structure with increasing ion fluence has been probed using nuclear magnetic resonance (NMR) and transmission electron microscopy (TEM), and associated with the corresponding changes of the magnetic behavior. A decrease of both the structural order and the moment in saturation is observed. Specifically, we detect a direct transition from a highly $L2_1$-ordered to a fully $A2$-disordered structure type and quantify the evolution of the $A2$ structural contribution as a function of ion fluence. Complementary TEM analysis reveals a spatially-resolved distribution of the $L2_1$ and $A2$ phases showing that the $A2$ disorder starts at the upper part of the films. The structural degradation in turn leads to a decreasing magnetic moment in saturation in response to the increasing fluence. 

\end{abstract}


\end{frontmatter}


\section{Introduction}

Highly $L2_1$-ordered Heusler compounds (X$_2$YZ) are promising candidates for half-metallic ferromagnetism \cite{deGroot1983,Kandpal2007,Wurmehl2006,Yang2012,Sakuraba2014,Sakuraba2014b} whose 100\,$\%$ spin polarization gives rise to interesting applications in spintronic devices, such as magnetic tunnel junctions (MTJ) or giant magnetoresistance (GMR) devices \cite{Schmalhorst2004,Sakuraba2006,Ishikawa2006, Sakuraba2006b,Sakuraba2010}. In combination with its high Curie temperature (985\,K), saturation moment (5\,$\mu_B$) and large band gap for the minority spin states (0.4-0.8\,eV) \cite{Kandpal2007, Sakuraba2006, Ishikawa2006, Brown2000, Picozzi2002}, \cms\ is especially well-suited for such applications. 
Chemical order has strong impact on the properties in general. This relationship is in particular important for half-metallic ferromagnetic materials, since their spin polarisation is downgraded rapidly for certain types of disorder \cite{Picozzi2004, Miura2004, Raphael2002, Kogachi2009}. As a consequence, the performance of any application based on the spin polarisation, e.g. GMR devices, will in turn strongly depend on the degree of chemical order of the material \cite{Wurmehl2016b}.
Hence, a control of the structure-property relationship is a crucial factor for their use in high-end applications. 
The classical route to address and control the structural order and, thus, the properties of such highly functional materials is post-growth annealing, where the annealing temperature and/or annealing time have to be optimized. 

Another route to control the structure and magnetic properties of functional materials is the irradiation with light ions, which is known to strongly modify their magnetic properties 
\cite{Fassbender2014, Bali2014, Heidarian2015, Ravelosona2000}. 
Studies on \cms\ thin films reported improved structural properties upon the irradiation with \He\ ions within a certain range of fluences ($1\times$10$^{14}$ - $5\times$10$^{14}$\,ions/cm$^2$, 30\,keV), going along with an improvement of the electronic and magnetic properties \cite{Gaier2009}. However, the changes in the observed saturation moment were on the order of a few percent while the structural results were extracted from small changes in the ratio of two superlattice reflections (200 and 400), whose absolute intensities were small, such that changes may be easily over/-underestimated.

In this work, we aim to track irradiation-induced chan-ges of the crystallographic order at the local level. To ensure  full characterization of the structural order, we applied both local (TEM) as well as volume-integral probing (NMR) of \He -irradiated \cms\ thin films.
The combination of magnetometry, nuclear magnetic resonance (NMR) and transmission electron microscopy (TEM) allows to get insights into the structure-property relationship, in particular linking both local and macroscopic properties. 
NMR probes the local environment of the investigated nuclei over the entire sample volume. It has already been shown that NMR is a potential method to resolve the evolution of structural order in functional materials \cite{Wurmehl2016b,NBH79,Endo1992,Jay1996,pan97,Inomata2006,KMW06,Inomata2008,Wurmehl2011,Wojcik2012,Rodan2013,Wurmehl2014,Gellesch2017}. By means of \co\ NMR, we observe a degradation of the crystallographic order on a local scale from $L2_1$ to $A2$ order upon increasing \He\ ion fluence. We were able to deduce the amount of the respective $L2_1$- and $A2$-ordered phases for each applied fluence.  
We complement the NMR results by TEM which provides structural information of selected regions of the sample. 
The evolution of the structure is linked to the macroscopic magnetic properties showing a decrease of the saturated moment in response to the ion irradiation.\\

\section{Sample details and experimental details}

\cms\ thin films of 40\,nm thickness covered with a 5\,nm thick Ta layer were deposited epitaxially on a (100) MgO substrate by magnetron sputtering and annealed at 500\,$^\circ$C. Samples of similar sizes ($\sim$ $7\times7$\,mm$^2$) have been cut and have been exposed to a 15\,keV \He\ ion beam with fluences between $1\times$10$^{13}$ and $5\times$10$^{15}$ ions/cm$^2$.  The ion beam was incident perpendicular to the film surface. This irradiation has been carried out at ambient temperature at the Ion Beam Center at the Helmholtz-Zentrum Dresden-Rossendorf. 
The displacements per atom (dpa) have been simulated using the binary collision approximation (Stopping and Range of Ions in Matter (SRIM) package) \cite{Ziegler2010}. According to these simulations, 15\,keV \He\ ions provide a nearly flat distribution of atomic displacements within the 40\,nm thick \cms\ film, while causing negligible intermixing with the capping layer and MgO substrate. The simulated dpa within the \cms\ film varied from 0.001 to 0.56 for fluences of $1\times$10$^{13}$ to $5\times$10$^{15}$ ions/cm$^2$, respectively (see Fig.~2 of supplementary material), leading to a statistical variation of the local ordering. NMR is an ideal tool to investigate such statistical variations.\\
The magnetization has been measured at 300\,K using a superconducting quantum interference device (SQUID) from Quantum Design in the field range between -20 to 20\,kOe. Please note, that the magnetization measured at 300\,K does not significantly differ from the respective 5\,K/0\,K values due to the high Curie temperature of 985\,K \cite{Ritchie2003}.
Zero-field (ZF) NMR measurements were performed at 5\,K with an automated, coherent, phase-sensitive and frequency-tuned spin-echo spectrometer (NMR Service, Erfurt, Germany). A solid-echo sequence, consisting of two 90$^{\circ}$ pulses of 0.6\,$\mu$s width, separated by 5\,$\mu$s was used to observe the spin echo. This pulse sequence was repeated up to 2000 times to get a satisfying signal-to-noise ratio, while the repetition time between consecutive pulse sequences was kept long enough (100\,ms) to prevent spin-lattice-relaxation effects on the measured spin-echo. The recorded spin-echo was integrated over the measured frequency range, which was swept in 1\,MHz steps. The obtained NMR spectra were corrected for the enhancement factor as well as for the $\nu^2$-dependence, resulting in relative intensities which are proportional to the number of nuclei with a given NMR resonance frequency.\\
High-resolution transmission electron microscopy (HRTEM) analysis was performed using an image C$_{\text s}$-corrected Titan 80-300 microscope (FEI) operated at an accelerating voltage of 300\,kV. High-angle annular dark-field scanning transmission electron microscopy (HAADF-STEM) imaging and element mapping based on energy-dispersive X-ray spectroscopy (EDXS) were performed at 200\,kV with a Talos F200X microscope (FEI) equipped with an X-FEG electron source and a Super-X EDXS detector system. Prior to TEM analysis, the specimen mounted in a double-tilt analytical holder was placed for 10\,s into a Model 1020 Plasma Cleaner (Fischione) to remove contamination. 
Classical TEM specimen preparation based on sawing, grinding, polishing, dimpling, and final Ar ion milling was applied in the case of the sample irradiated with $5\times10^{14}$\,ions/cm$^2$.

\section{Experimental Results}

\subsection{Magnetometry}

\begin{figure}[t]
\centering
\includegraphics[width=\linewidth]{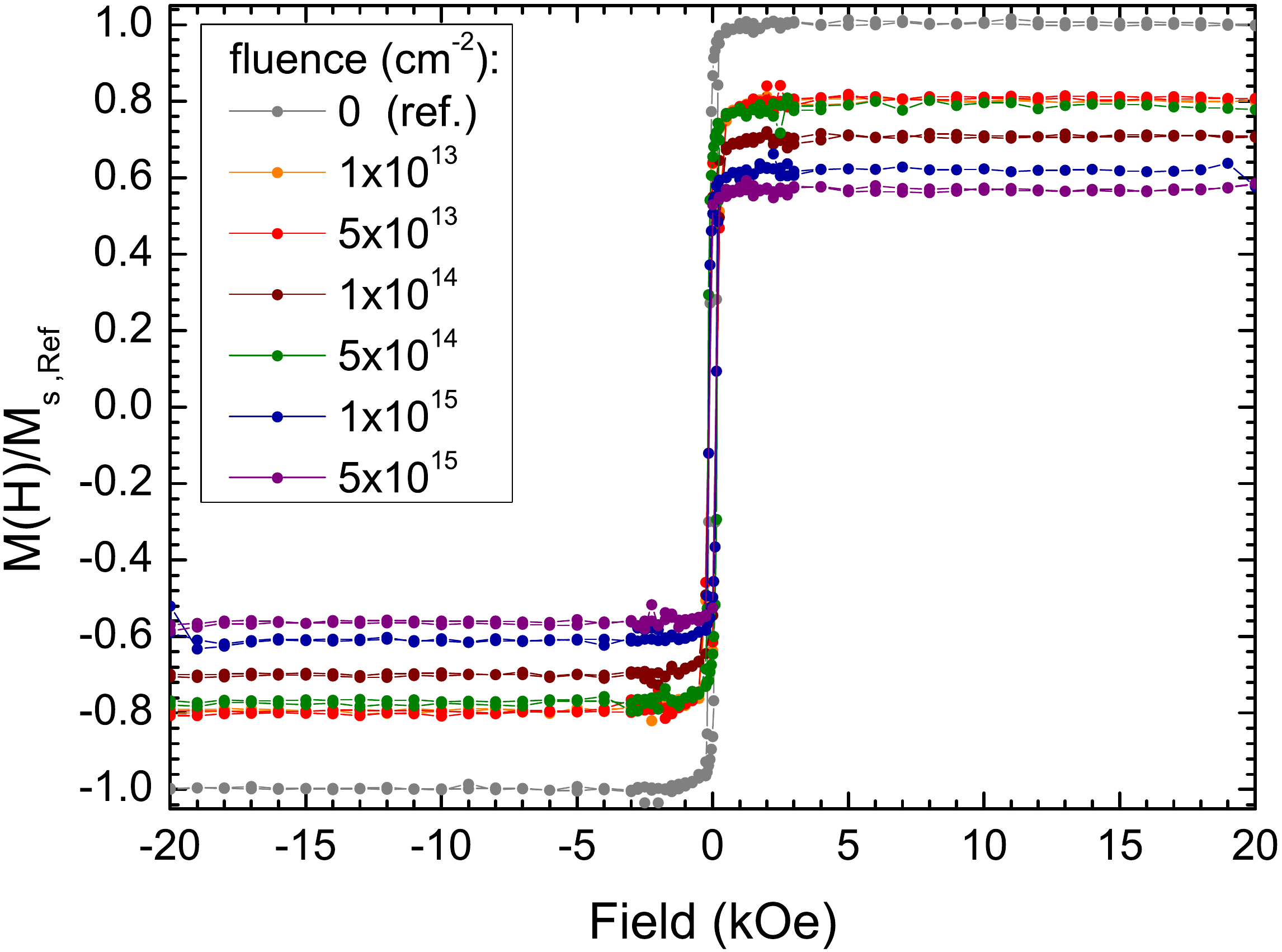}
\caption{\label{fig:Magn} (Color online) Magnetization of \cms\ thin films at 300\,K in fields up to 20\,kOe for a non-irradiated reference sample (grey dots and lines) and the irradiated samples with \He\ ion fluences ranging from 10$^{13}$\,ions/cm$^2$ up to $5\times10^{15}$\,ions/cm$^2$. The data have been normalized to the data of the reference sample in saturation (20\,kOe). The diamagnetic background from the substrate and the capping has been subtracted (See text for details.)}
\end{figure}

\begin{figure*}[t]
  \centering
   {\includegraphics[width=\textwidth]{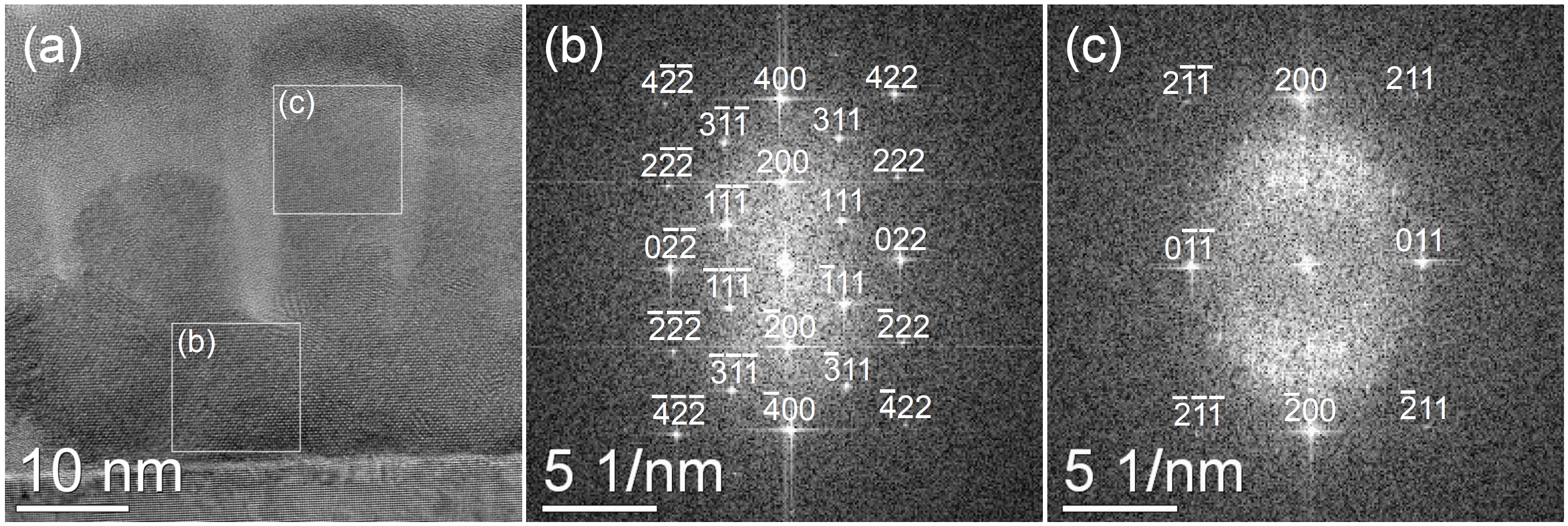}  }
 \caption{Cross-sectional HRTEM image (a) and corresponding fast Fourier transforms for the lower part (b) and upper part (c) of the sample irradiated with $5\times 10^{14}$\,ions/cm$^{2}$, as obtained from the marked square areas.}
  \label{fig:TEM-FT}
\end{figure*}

The results from SQUID magnetometry performed at 300\,K of the irradiated \cms\ thin films are displayed in Fig.~\ref{fig:Magn} together with the magnetization loop of a non-irradiated reference sample from the same batch. The diamagnetic background from the substrate and the Ta capping, which leads to a linear decrease (increase) at high (low) field values, has been subtracted. While the coercive field stays constant within error bars ($\sim 125 - 160$\,Oe), the saturation magnetization decreases noticeably  upon increasing the fluence. The biggest step occurs between the non-irradiated reference sample and the sample irradiated with the lowest fluence applied (10$^{13}$\,ions/cm$^2$, 20$\%$ reduction). The sample irradiated with the highest fluence ($5\times 10^{15}$\,ions/cm$^2$) exhibits a nearly halved saturation moment.   
Since magnetic properties of Heusler compounds are strongly linked to the underlying structural order, in
the following, we will monitor the structural changes upon irradiation using TEM and NMR.

\subsection{HRTEM}

To study the evolution of the microstructure, HRTEM imaging of the non-irradiated reference and the sample irradiated with a fluence of $5\times 10^{14}$\,ions/cm$^{2}$ were performed. In the non-irradiated state, fast Fourier transformation (FFT) analysis (see Fig.~4 of supplementary material) points to the presence of the $L2_1$ structure, in good agreement to our NMR results (see following section). Regarding the sample irradiated with $5\times 10^{14}$\,ions/cm$^{2}$, TEM analysis can differentiate between two sample regions. Evaluating the area close to the MgO substrate, i.e. the lower part of the \cms\ film, the diffractograms obtained by FFT are indexed based on the $L2_1$ structure in [101] zone axis geometry [Fig.~\ref{fig:TEM-FT}(b)], i.e. the same way as for the reference sample. In some parts of the upper \cms\ film region, however, particular $L2_1$ reflections are strongly decreased in intensity or missing even completely, pointing to an increased disorder. These diffractograms can be better described with the $A2$ structure, also in [101] zone axis geometry [Fig.~\ref{fig:TEM-FT}(c)]. These findings agree well with the NMR results, which found roughly a 50:50 distribution of $L2_1$ and $A2$ structure in this sample. While the NMR measurements cannot determine the location of these structure types within the \cms\ layer, HRTEM analysis, showed that $A2$ disorder is mostly present in the upper part of the film. Please note that light grey parts of Fig.~\ref{fig:TEM-FT}(a) indicate the presence of  oxide columns growing from the surface deep into the film (see supplementary material for details). These inclusions do not interfere with magnetization and NMR measurements, as discussed in detail in section~\ref{Disc}.

\subsection{NMR}

\begin{figure}[t]
\centering
\includegraphics[width=0.9\linewidth]{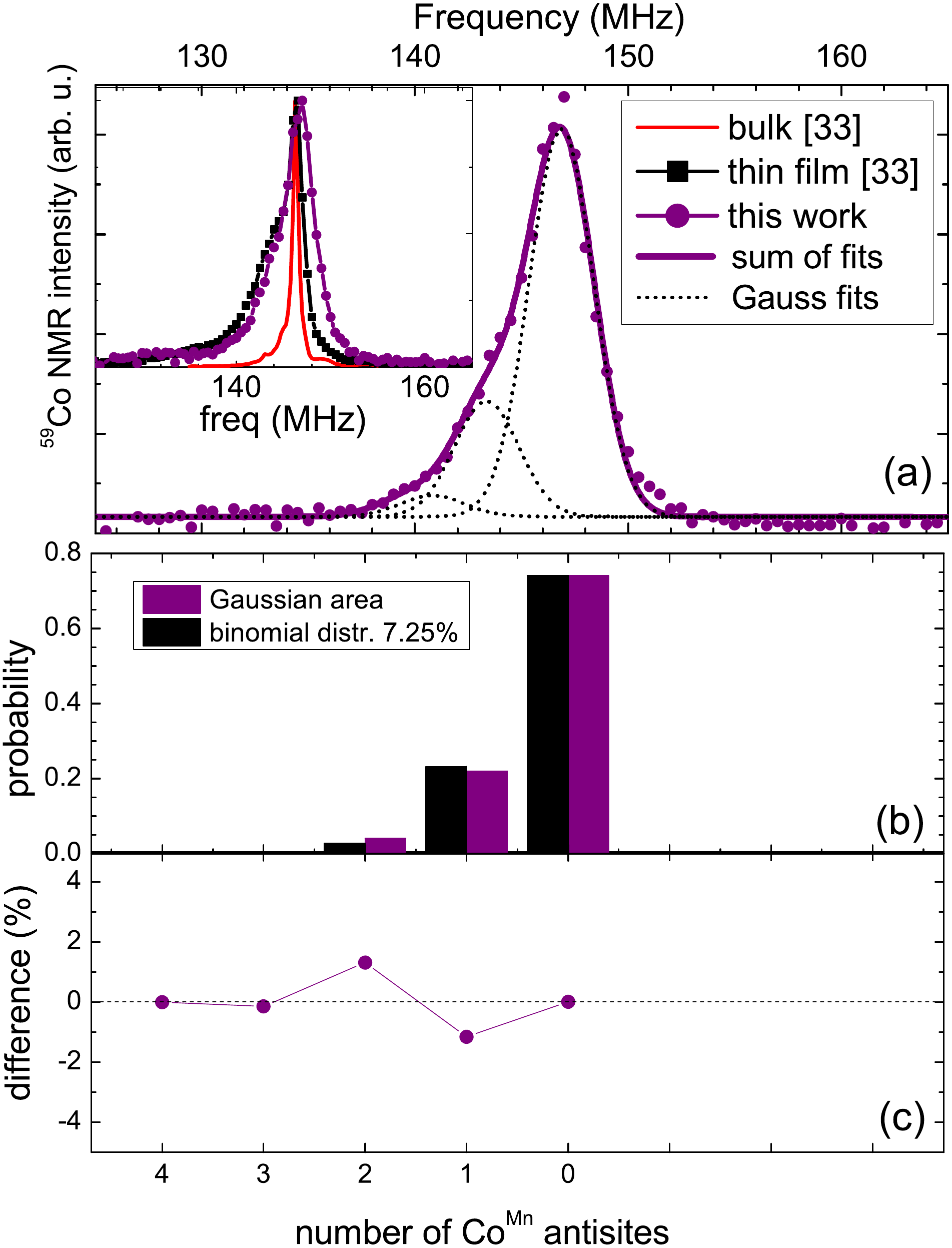}
\caption{\label{fig:spec0} (Color online) (a) \co\ NMR spectra at 5\,K for the non-irradiated reference sample (purple dots). Dotted lines are fits with Gaussian lines, representing the off-stoichiometry. The thick purple line is the sum of these Gaussian lines. The inset shows a comparison of the \co\ NMR spectra of the reference sample (purple dots and line), to a \cms\ thin film annealed at 550$^\circ$C (black squares and line) and a bulk polycrystalline \cms\ sample (red line) \cite{Rodan2013}. (b) Probability $P(n,x)$ of the binomial distribution function for $x=7.25$\,\% (black columns) in comparison to the relative areas of the Gaussian fits to the spectrum (purple columns). (c) Difference between the relative areas of the Gaussian lines and the binomial probability. }
\end{figure}

The inset of Fig.~\ref{fig:spec0}(a) shows the \co\ NMR spectrum of the non-irradiated reference sample (annealed at 500$^\circ$C) in comparison to \co\ NMR spectra of an $L2_1$-ordered polycrystalline bulk sample and a \cms\ thin film, annealed at 550$^\circ$C \cite{Rodan2013}. Similar to these samples, the spectrum of the non-irradiated reference sample shows a maximum at around 147\,MHz, characteristic for the \cms\ phase, and a well-defined, reasonably small linewidth, representing a long-ranged, truely $L2_1$-ordered \co\ environment (4\,Mn and 4\,Si atoms as nearest neighbors).  

\begin{figure}[t]
\centering
\includegraphics[width=\linewidth]{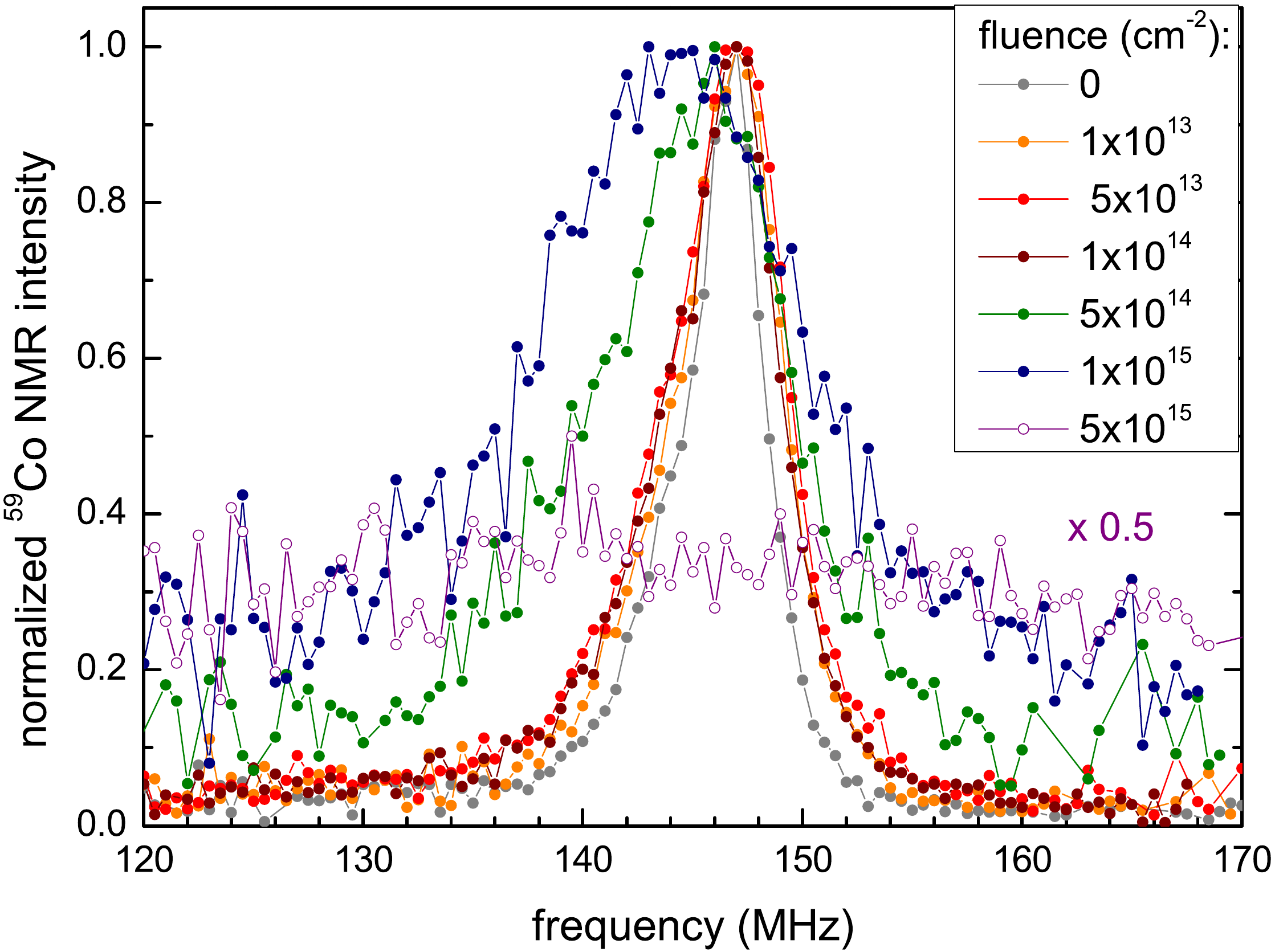}
\caption{\label{fig:spec-norm} (Color online) Normalized \co\ NMR spectra of the irradiated films in comparison to the non-irradiated reference sample.}
\end{figure} 

It exhibits a low frequency shoulder, stemming from a slight off-stoichiometry of excess Co atoms occupying Mn sites (Co$^{\text{Mn}}$ antisites) \cite{Rodan2013}. The amount of Co$^{\text{Mn}}$ antisites can be estimated by fitting the spectrum with Gaussian lines and comparing their relative areas to a binomial distribution function, resulting from a random atom model for the first shell environment of \co\ \cite{Rodan2013, Wurmehl2007,  Schaf1983, Wurmehl2009}:

\begin{equation}
P(n,x) = \frac{N!}{(N-n)!n!}(1-x)^{N-n}x^n  \, ,
\end{equation}

where $P(n,x)$ is the probability that $n$ sites out of $N$ sites are atoms of concentration $x$ on those $N$ sites. For the first Mn coordination shell (nearest neighboring Mn) of \co, $N$ is four and $n$ runs from 0 to 4 in integer numbers. Fig.~\ref{fig:spec0} shows such a fit with three equidistant Gaussian lines of equal width (a) and a comparison between their relative areas with a probability function $P(n,x)$ for $x=7.25$\,\% (b) for the non-irradiated reference sample. The agreement between the fits and the distribution function is very high, which is further proven in a plot of their difference (c), not exceeding 2\,\%.

Fig.~\ref{fig:spec-norm} shows the evolution of the corresponding \co\ NMR spectra upon the irradiation with \He\ ions for all applied fluences. 
Already for moderate fluences ($10^{13}$ - $10^{14}$\,cm$^{-2}$), the spectrum broadens slightly, compared to the non-irradiated sample, reflecting an increasing structural disorder. 
For higher fluences ($5\times 10^{14}$ - $10^{15}$\,cm$^{-2}$), the broadening gets even more pronounced and the peak frequency shifts to lower values. The spectrum of the sample irradiated with the highest fluence ($5\times 10^{15}$\,cm$^{-2}$) is completely smeared out over the whole measured frequency range and no distinct peak can be identified. Such a broad NMR resonance line is typical for a completely disordered $A2$ structure type, where, due to a complete intermixing of all atomic sites, a plethora of different first shell environments exist for the \co\ nuclei, leading to a distribution of many different hyperfine fields \cite{Inomata2006, Wurmehl2008} (see Fig.~1 of supplementary material for the $L2_1$ and $A2$ structure types, their corresponding first shell environments and expected NMR spectra). 

\begin{figure*}[]
  \centering
  \begin{minipage}[t]{0.32\textwidth}
    \label{nmr_all1}
    { \includegraphics[width=0.97\textwidth, clip]{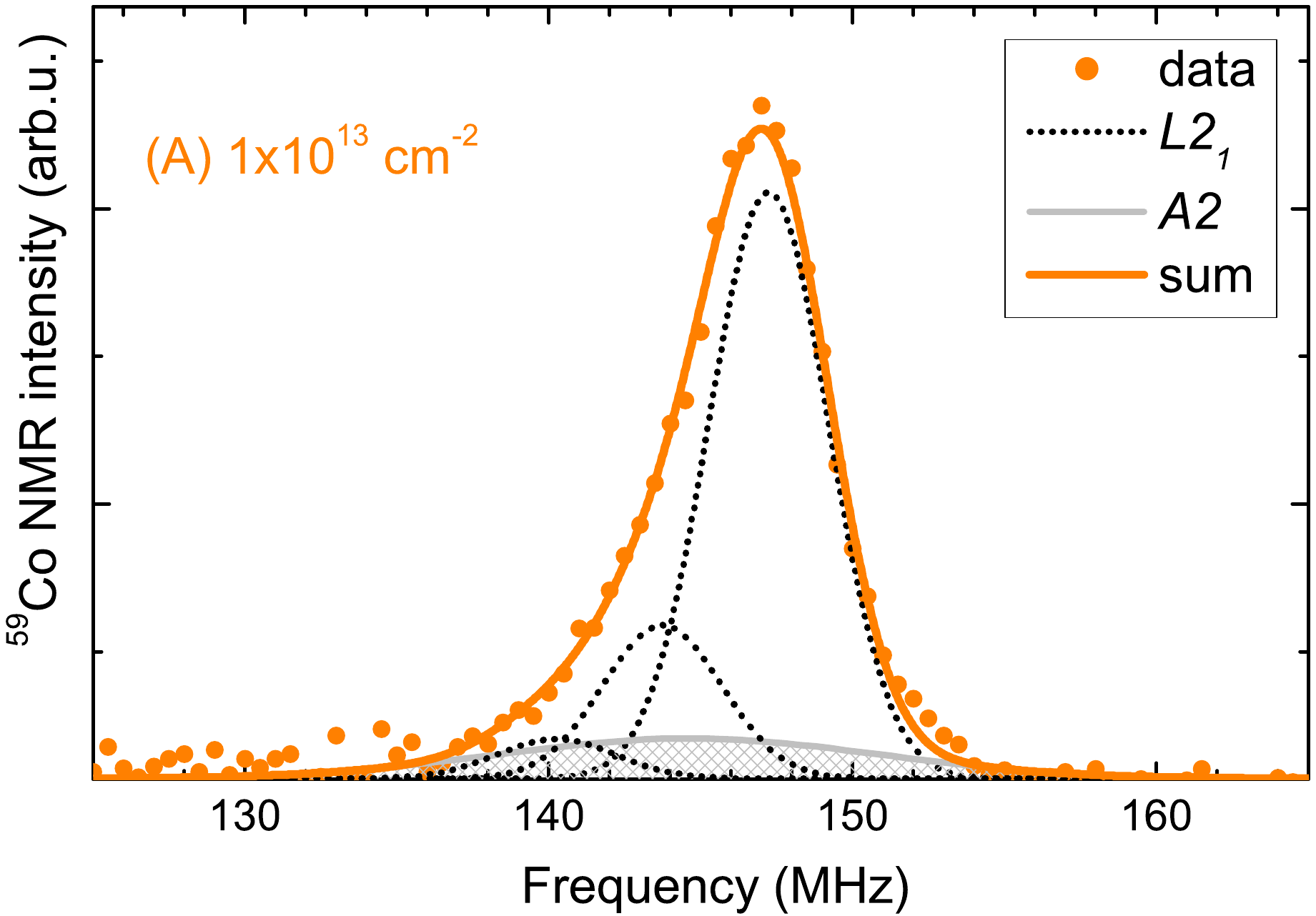}
  }
  \end{minipage}
  \begin{minipage}[t]{0.32\textwidth}
   \label{nmr_all2}
   { \includegraphics[width=0.97\textwidth, clip]{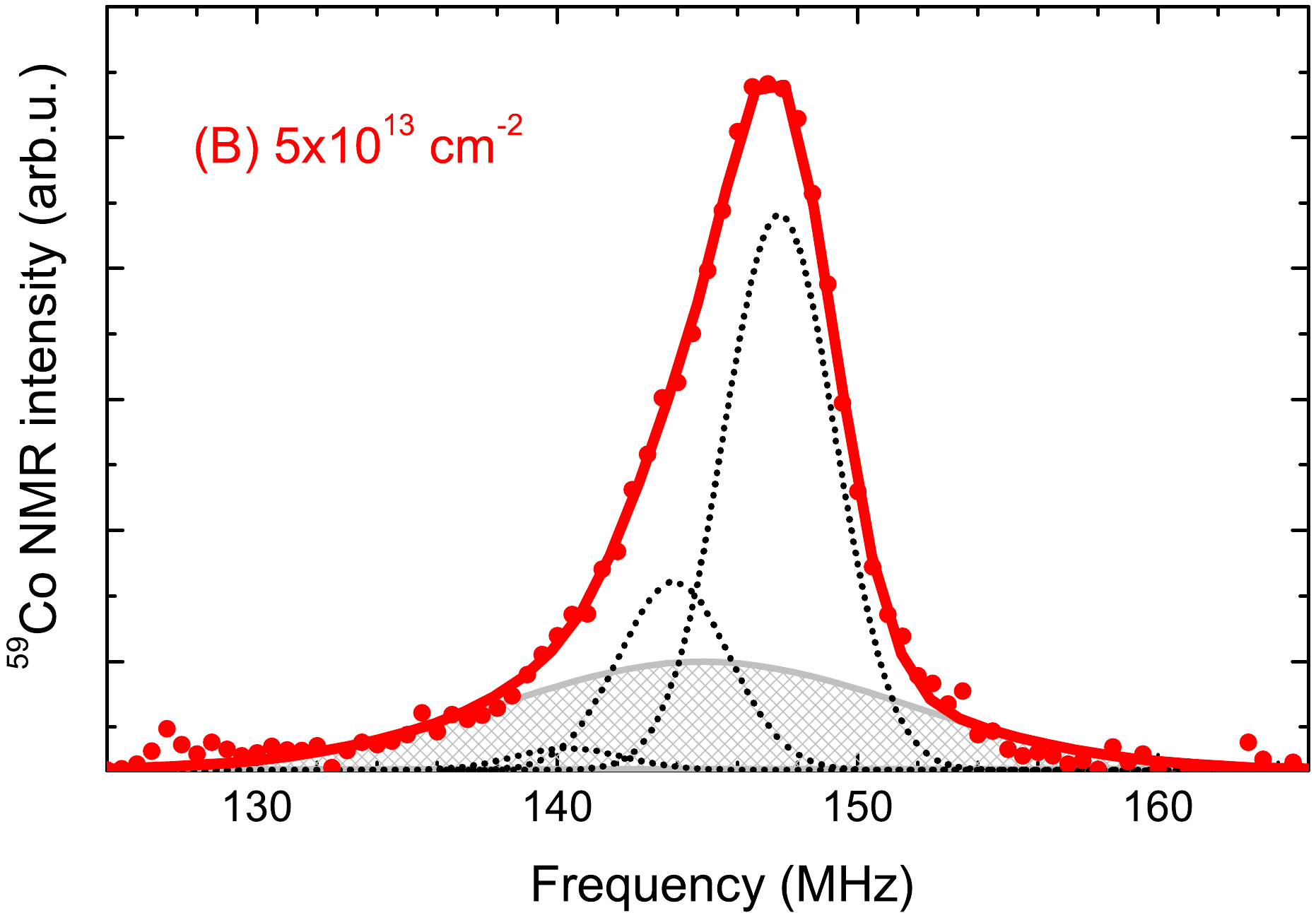}
  }
  \end{minipage}
   \begin{minipage}[t]{0.32\textwidth}
    \label{nmr_all3}
   { \includegraphics[width=0.97\textwidth, clip]{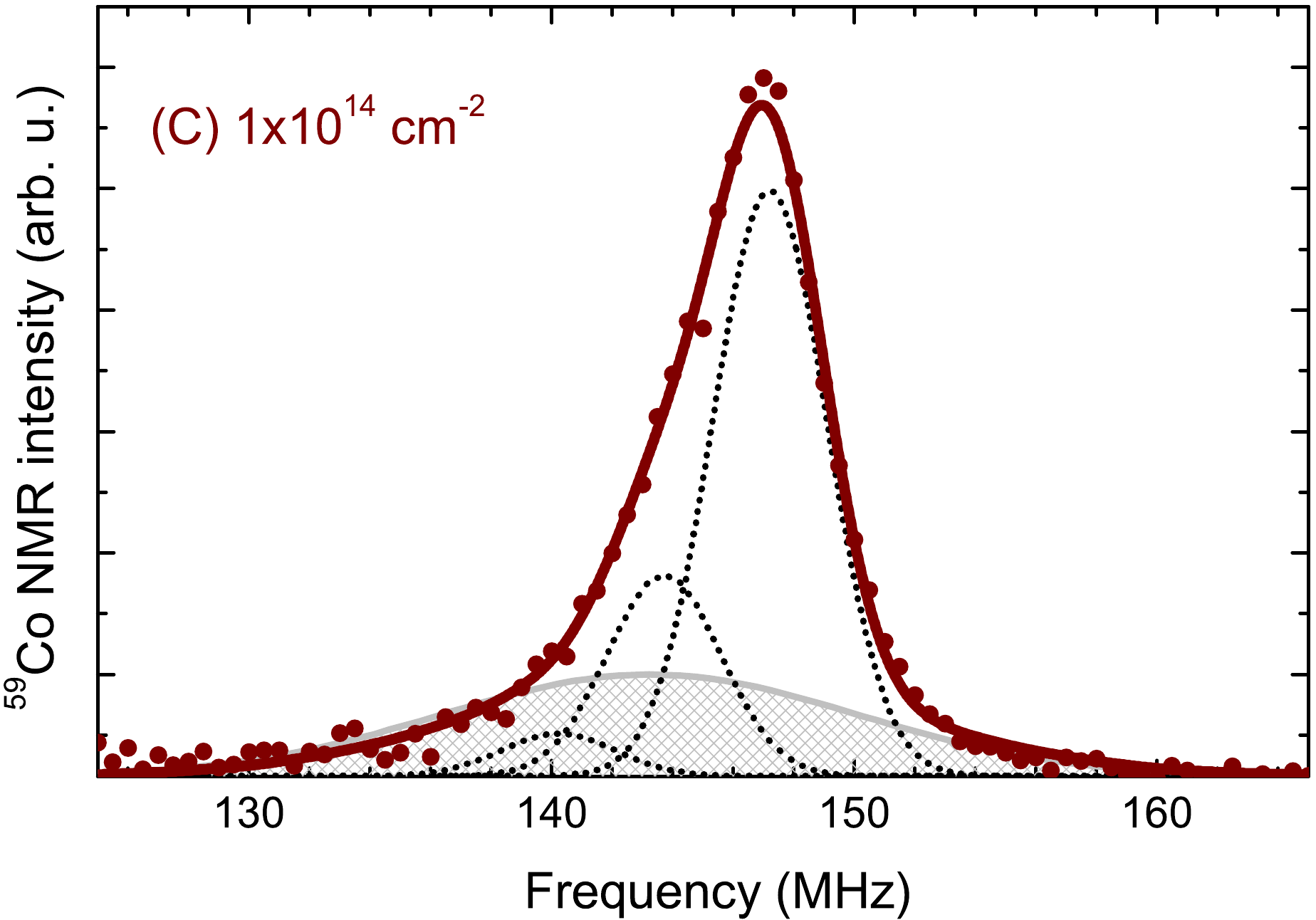}     	 
  }
  \end{minipage}
  \begin{minipage}[t]{0.32\textwidth}
    \label{nmr_all4}
    { \includegraphics[width=0.97\textwidth, clip]{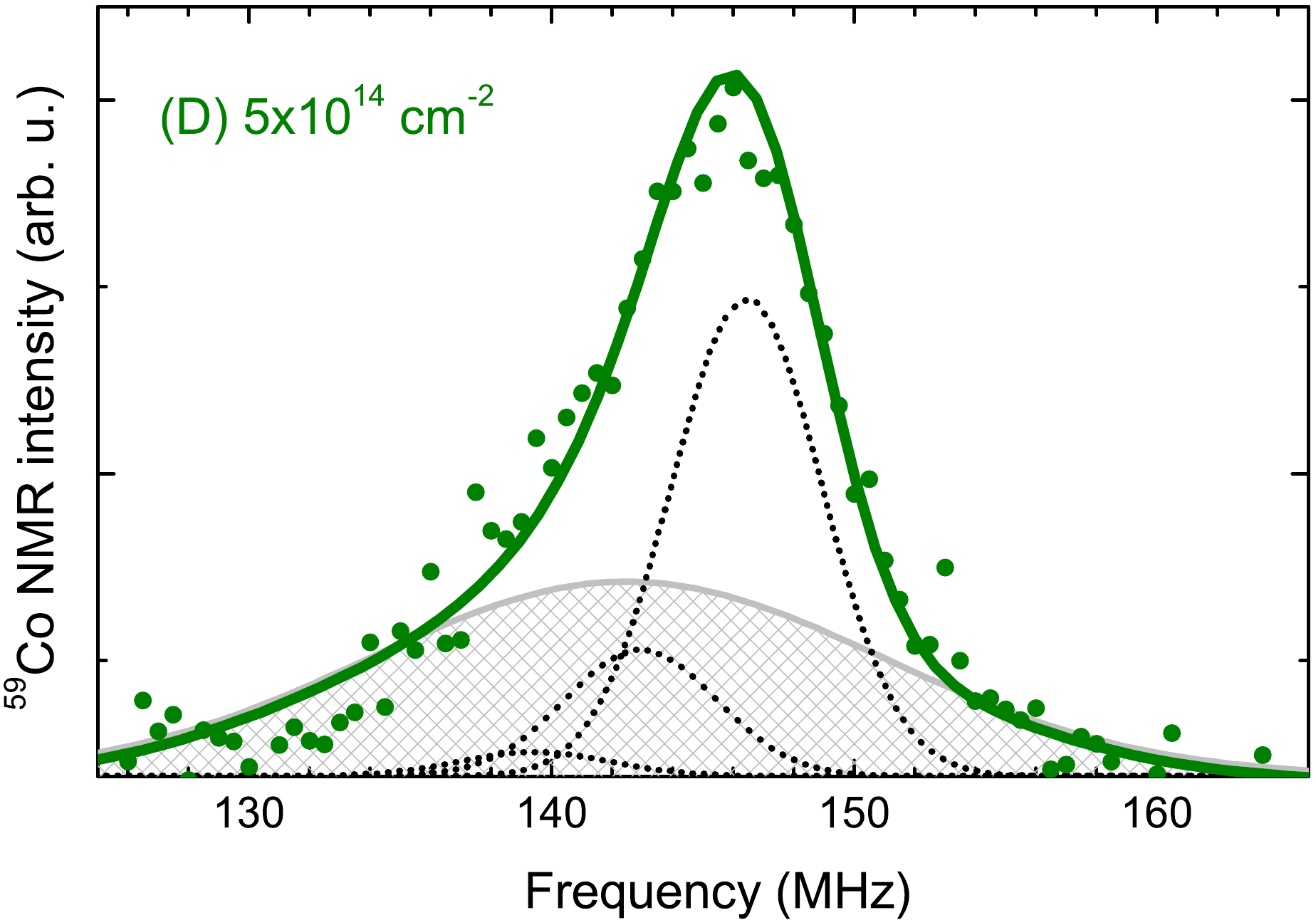}
  }
  \end{minipage}
  \begin{minipage}[t]{0.32\textwidth}
   \label{nmr_all5}
   { \includegraphics[width=0.97\textwidth, clip]{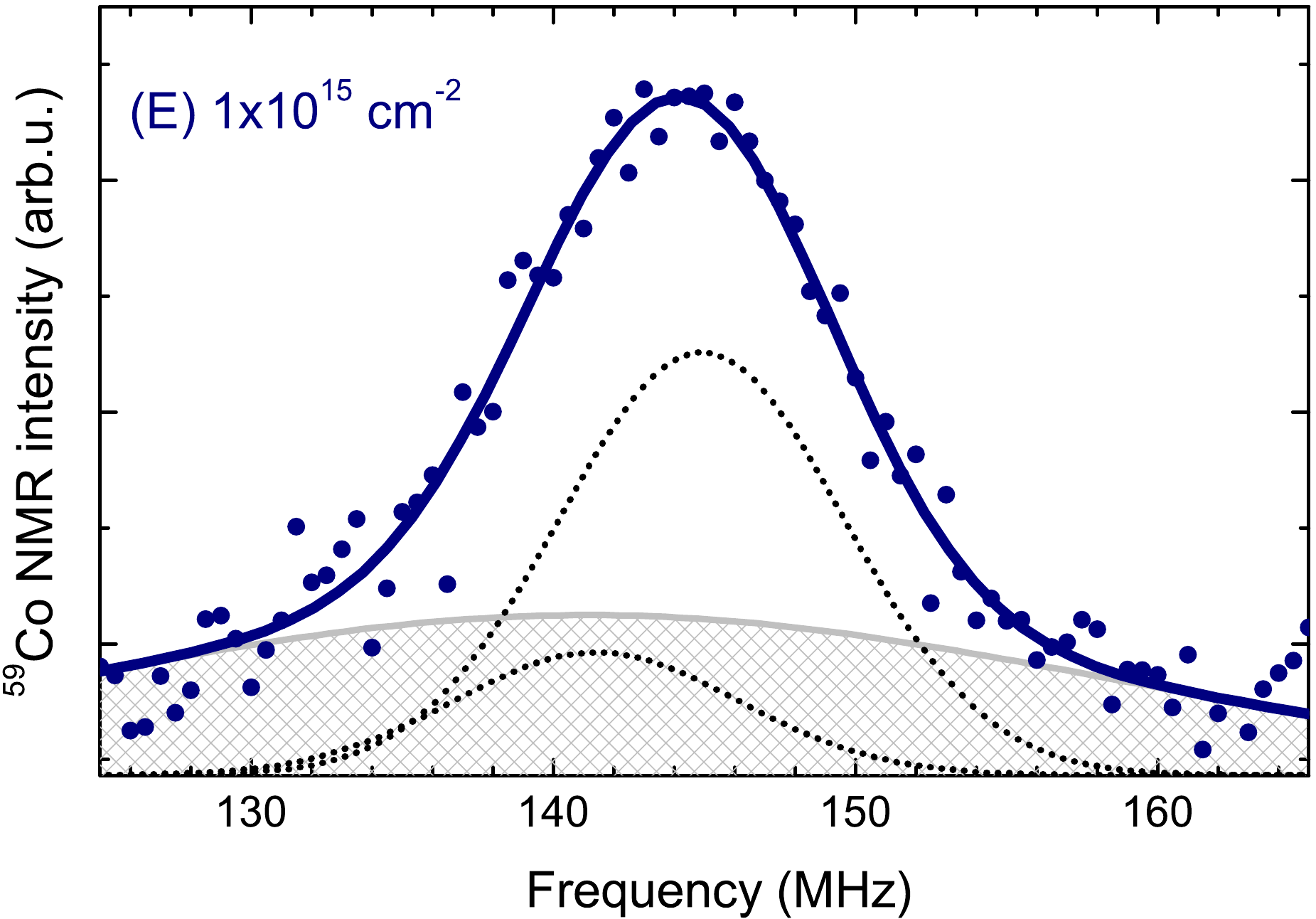}
  }
  \end{minipage}
    \begin{minipage}[t]{0.32\textwidth}
    \label{nmr_all6}
   { \includegraphics[width=1.05\textwidth, clip]{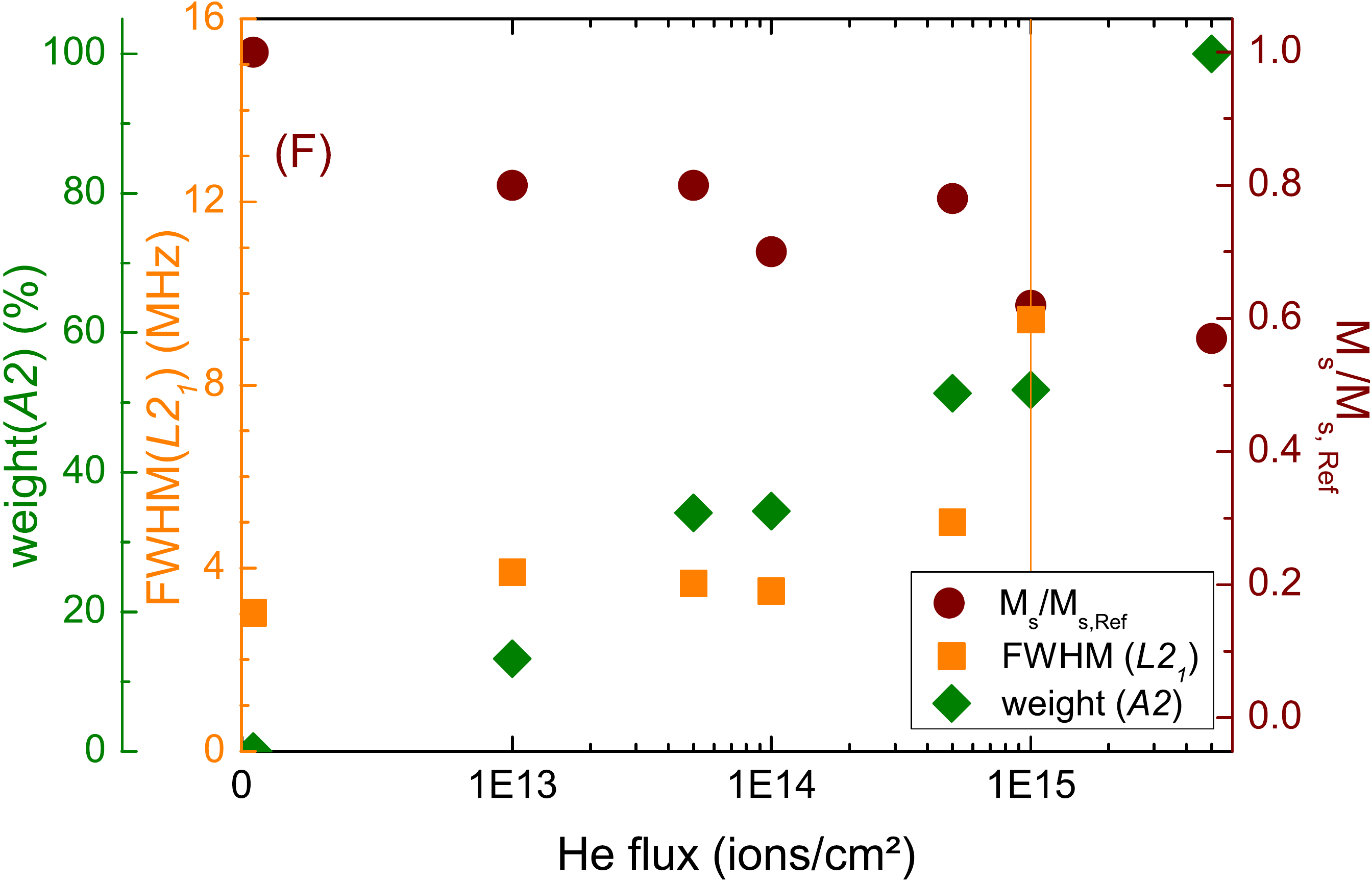}    	 
  }
  \end{minipage}
  \caption{(Color online) (A)-(E) \co\ NMR spectra at 5\,K for irradiated samples starting with the lowest flux of $1\times10^{13}$\,ions/cm$^2$ up to $1\times10^{15}$\,ions/cm$^2$. The corresponding fluence is written in the graphs. Lines are fits with Gaussian lines to the spectra. Dotted black lines denote the Gaussians representing the off-stoichiometric {\it L2}$_1$ ordered regions of the samples, while the filled light grey line corresponds to the {\it A2}-ordered regions of the sample. The thick line in the color of the respective data dots is the sum of all Gaussian lines. (F) Irradiation-dependent weight of the $A2$ structure type (dark green diamonds), FWHM of the $L2_1$ phase (orange squares), and the corresponding change in the saturation moment (dark red dots).}
  \label{fig:spec}
\end{figure*}

Figs.~\ref{fig:spec}(A)-(E) show Gaussian fits to the spectra of the irradiated samples. Three dotted lines indicate the off-stoichiometric $L2_1$ phase, similar as already discussed for the non-irradiated reference sample [Fig.~\ref{fig:spec0}(a)]. 
An additional broad Gaussian line (filled light grey line) represents the emerging $A2$ structure formation. This $A2$ contribution evolves already for the smallest fluence used. Its percentage increases upon increasing the fluence [see Fig.~\ref{fig:spec}(F)], until reaching 100\,\% for the sample irradiated with the highest fluence ($5\times 10^{15}$\,cm$^{-2}$, see Fig.~\ref{fig:spec-norm}). 
The full width at half maximum (FWHM) of the $L2_1$ peaks also increases [see Fig.~\ref{fig:spec-norm} and Fig.~\ref{fig:spec}(F)].
Due to the increased linewidths, only two Gaussians can be fitted for the $L2_1$ order in the last spectrum [Fig.~\ref{fig:spec}(E)]. The spectrum for the highest fluence ($5\times 10^{15}$\,cm$^{-2}$) is omitted in this plot, since its broad NMR line, ranging over the whole measured frequency range, represents a 100\,\% disordered $A2$ structure type.

Our NMR measurements show that for the considered range of \He\ fluences, ion irradiation leads to an increase of structural disorder at the local scale, going along with a strong impact on the macroscopic magnetic properties, even for the lowest fluences applied. 
We observe a direct crossover from the highly-ordered $L2_1$ to the completely disordered $A2$ structure upon increasing the \He\ fluence. 

\section{Discussion}
\label{Disc}

Several conclusions can be drawn from the NMR data:

The NMR spectra of all samples show a low-frequency shoulder, which can be fitted assuming three Gaussian lines, whose intensity distributions reflect a slight off-stoi-chiometry of 7.25\,\% excess Co on Mn sites. This kind of off-stoichiometry of an otherwise highly-ordered $L2_1$ Heusler structure has been frequently observed in thin films and is typical for stoichiometric sputter targets. It arises from differences in the sputtering rates of individual elements \cite{Sakuraba2006,Rodan2013,Wurmehl2009,Oogane2006,Inomata2008}. For the investigated samples, the antisite concentration $x$ has been found to be the same throughout the whole irradiation series (see supplementary material).\\

The $L2_1$ lines broaden with increasing ion fluence and an additional, very broad $A2$ line evolves.
These observations imply changes in the local structure of the \co\ atoms on different coordination shells: (i)
The probability to have a random distribution of all atoms in higher coordination shells increases. This disorder 
manifests mainly in the increasing FWHM of the $L2_1$ Gaussian lines (dark grey lines in Fig.\ref{fig:spec}) [(see Fig.~\ref{fig:spec}(F)]. (ii) At the same time, the probability to change also the first coordination shell, which for the non-irradiated sample consists of 4 Mn and 4 Si atoms, increases with increasing fluence. Since within the $A2$ type (dis)order the total number of different possible atomic configurations in all atomic coordination shells is very large, as is the number of their internal fields, these structural changes lead to the observed additional broad NMR line. For the highest fluences, the majority of Co atoms has a local environment different to those from the non-irradiated sample.\\

Regarding the type of emerging disorder upon ion irradiation, in principle other structure types such as $B2$ or $DO_3$ are imaginable. In fact, a similar study on 40\,nm thin films of \cms\ (originally $L2_1$-ordered with some small grains of $B2$ order) irradiated with \He\ ions of much higher energy (150\,keV, 10$^{15}$ - 10$^{16}$\,ions/cm$^2$) reported an increase of $B2$ and $DO_3$ disorder upon increasing fluence by means of XRD and HAADF-STEM \cite{Abdallah2016}. However, our NMR spectra do not show any sign of these structure types. For $B2$, with an intermixing of Mn and Si sites, the NMR spectrum should show additional, well-separated peaks with a fixed spacing on both sides of the central peak \cite{Inomata2006,Wurmehl2008}. In the case of $DO_3$ order (Co-Mn intermixing), peaks with different intensities and spacings on both sides of the central peak would be expected, stemming from the $1^{\rm{st}}$ and $2^{\rm{nd}}$ coordination shells \cite{Kandpal2007}. The broad additional line in the NMR spectrum of our samples, which evolves upon irradiation, indicates emerging $A2$ disorder, in excellent agreement with the HRTEM results.\\

\begin{table*}[t]
\centering
\caption{\label{table} Overview of studies on \He -irradiated \cms\ thin films, their corresponding experimental conditions and the resulting change of structural properties.}
{
\begin{tabular}{lcccr}
 & & & & \\
Reference & E  & fluence  & structure of & structural evolution\\
  & (keV) & (ions/cm$^2$) & starting material& \\ \hline 
  & & & & \\
this work  & 15 & 10$^{13}$ - 10$^{15}$ & $L2_1$ &  decrease of structural order ($A_2$)  \\
Gaier {\it et al.} \cite{Gaier2009}  & 30& 10$^{14}$ - 10$^{16}$ & $B2$ + $A2$ & increase of structural order ($B_2$) \\ 
Abdallah {\it et al.} \cite{Abdallah2016}  & 150 & 10$^{15}$ - 10$^{16}$ & $L2_1$ + $B2$  grains & decrease of structural order ($B_2$ and $D0_3$) \\ 
\end{tabular}
}
\end{table*}

With this study, there are now three different reports on the impact of \He\ irradiation on \cms\ thin films (which are summarized in table~\ref{table}): Gaier {\it et al.} reported an improvement of structural and magnetic properties upon the irradiation with 30\,keV \He\ ions for a certain range of fluences (around 10$^{14}$ ions/cm$^2$)  \cite{Gaier2009}; a more recent study showed an increase of $B2$ and $DO_3$ disorder and a reduced magnetization by using 150\,keV \He\ ions of 10$^{15}$ to 10$^{16}$ ions/cm$^2$ \cite{Abdallah2016, Abdallah2017}; while the present work reports a direct crossover from highly-ordered $L2_1$ to completely disordered $A2$ structure (15\,keV, 10$^{13}$ - 10$^{15}$ ions/cm$^2$), without inducing other structure types. 
SRIM calculations (not shown) showed that there is only a slight difference in the average Co displacement upon changing the ion beam energy from 15 to 30\,keV \He\ ions. The irradiation with 150\,keV \He\ ions is supposed to lead to bigger differences from local heating due to the larger deposited energy.
Apart from the different energies, an explanation for the observed variations might be the different starting conditions. Gaier {\it et al.} used $B2$-ordered films with a certain admixture of $A2$-disordered regions as starting material. They conclude that the increase of $B2$ order originates from initially $B2$-ordered regions, due to Co - Mn and Co - Si exchanges caused by mobile vacancies introduced by the irradiation. Similarly, the $L2_1$-ordered films from Abdallah {\it et al.} already contained some very small grains of $B2$ order, which might have favored a (in this case degrading) $B2$ transformation starting from these initial nucleation sites. 
The starting material of the present study is highly $L2_1$-ordered. The displacement of atoms caused by the incident ion beam produces vacancies, which will stochastically recombine with thermally diffusing atoms resulting in disorder.  
Since there are no $B2$ nucleation sites, we expect an arbitrary intermixing of all atoms at all sites, leading to the completely disordered $A2$ structure, as observed here.\\
It should also be noted that the reported improvement of $B2$ structure and magnetic properties in Ref.~\cite{Gaier2009} is limited to a certain range of fluences (1 - 5 $\times$10$^{14}$\,ions/cm$^2$) for which only a very subtle increase of structural order and saturation moment were observed. 
For higher fluences also this study found a substantial degradation of structural and magnetic properties.\\

One of the advantages of the NMR technique is that it allows to probe the local environments of all Co atoms in the film sample in one measurement. (Assuming a relative permeability of 600 (Co) and taking the reported resistivity values of \cms\ at 5\,K, which range from 4 - 40\,$\mu\Omega$cm \cite{Sakuraba2010,Ritchie2003,Raphael2001}, the skin depth of the penetrating microwave amounts to 400 - 1000\,nm \cite{Abragam}, which is far more than the film thickness.) Since we are probing the full volume of the film without any losses due to film thickness, the NMR method does not reveal from which position in the samples a certain NMR signal is excited, e.g., from Co atoms at the top or at the bottom of the film.
Hence, we complemented the NMR results with TEM which can provide spatially resolved information about the local structure at a specific sample position. 

According to our HRTEM results, there is a depth-dependent structural order in the irradiated samples: regions next to the surface are mainly $A2$-ordered, while regions next to the substrate still retain the original $L2_1$ structure. In comparison to the SRIM simulations, which showed a rather flat distribution of atomic displacements within the \cms\ film (see Fig.~2 of supplementary material), this is rather surprising. One may consider this observation to be triggered by the presence of oxide columns growing from the surface deep into the film (down to about 20~nm deep into the film; see Fig.~5 of supplementary material and discussion in following paragraph). Here, oxygen may    
nucleate the vacancies leading to a higher probability for atomic displacements in their vicinity.\\
Our results underline that structure analysis by NMR, which allows quantitative determination of phase fractions on an integral scale, and by TEM, which enables imaging down to the atomic scale, complement each other very well. A similar combination of NMR and TEM was already successfully used to study the chemical composition of multi-element nanostructures \cite{Gellesch2017}.\\

We now turn to the relation between (local) structure and macroscopic magnetic properties.
Note that the absolute value of the saturated moment of our non-irradiated reference sample (2.7\,$\mu_B$/f.u. at 300\,K, which translates to 2.8\,$\mu_B$ at 5\,K \cite{Ritchie2003}) is lower than the theoretically expected and experimentally verified value of 5\,$\mu_B$/f.u. \cite{Brown2000,Raphael2001}. Two issues contribute to the lower absolute value of the saturated moment:
(i) Reduced saturated moments have also been reported for other $L2_1$-ordered Heusler thin films sputtered from stoichiometric targets, e.g., Co$_2$FeSi (4.5 - 5\,$\mu_B$ instead of 6\,$\mu_B$) \cite{Wurmehl2009, Tezuka2006, Schneider2006}. This difference most likely results from the mentioned off-stoichiometry due to selective sputtering. A polycrystalline bulk sample intentionally prepared with a corresponding off-stoichiometry, namely Co$_{2.1}$Mn$_{0.9}$Si, indeed showed a reduced saturation moment of 4.6\,$\mu_B$/f.u. (see Fig.~3 of supplementary material). Other theoretical and experimental work on \cms\ found comparable effects of Co antisites 
on the corresponding saturation moments (4.75\,$\mu_B$ - 4.9\,$\mu_B$, respectively) \cite{Picozzi2004, Raphael2002, Kogachi2009}.
However, the deviation from the expected value for our reference sample is too big (nearly 50\,\%) to be solely explained by a slight Co excess. (ii) In fact, our samples contain considerable amounts of oxide phases (see detailed discussion in supplementary material). In the case of magnetization data, these oxide phases do not contribute to the ferromagnetic signal, but lead to an overestimation of the intrinsic \cms\ mass when calculating the magnetization and, thus, to a reduced absolute saturation moment.
Importantly, the presence of these oxides does not interfere with the NMR analysis. Measurements of the variations of the local \co\ environment have been made from the same initial state on all samples, namely
from the remaining non-oxidized parts of the samples, which gave rise to the ferromagnetic SQUID signals. 
These are well-characterized \cms\ regions, as proven by HRTEM (Fig.~\ref{fig:TEM-FT}) and NMR analysis [see inset of Fig.~\ref{fig:spec0}(a)]. Both the off-stoichiometry and the overestimation of the intrinsic \cms\ mass due to the presence of oxides are expected to be present and on the same order of magnitude in all samples before the irradiation process.
Our discussion of the changes in the structure-property relationship of the \cms\ thin films upon the irradiation with \He\ ions is therefore straightforward and not affected by the presence of oxide inclusions or slight off-stoichiometry.\\ 

Finally, we see a clear trend of decreasing moment with increasing ion fluence mimicking the inverse trend of increasing disorder with increasing fluence.
These observations are in line with previous studies showing a strong dependence of the magnetic moment on the crystal structure \cite{Tezuka2006, Kudry2005, Wurmehl2006b}. Even for moderate fluences (around 10$^{14}$\,ions/cm$^2$), where, based on XMCD measurements, an increasing saturation moment has been reported for \cms\ thin films irradiated with \He\ ions of similar energy (30\,keV) \cite{Gaier2009}, we see a clear trend towards lower saturation moments and a substantial fraction of $A2$ structure. Hence, ion-irradiation can subtly vary the local environment, thereby revealing structure-property relationships in Heusler compounds.

\section{Conclusion}

Combined magnetization, NMR, and TEM measurements of \cms\ thin films under the impact of \He\ irradiation showed that, even for the lowest fluences applied, the moment in saturation decreases, going along with an increased structural disorder. The disorder manifests itself in a transition from a highly $L2_1$-ordered to a completely disordered $A2$ structure in a sample irradiated with the highest fluence applied ($5\times 10^{15}$\, \He\ ions/cm$^{2}$). The direct transition from  $L2_1$ to $A2$ has been revealed by NMR measurements, which could deduce the amounts of $L2_1$- and $A2$-ordered phases in each of the thin films. The NMR measurements also revealed a slight off-stoichiometry of 7.25\,\% Co excess, which is constant throughout the films, indicating that it originates from their preparation process.  
 Complementary HRTEM analyses added local information of the distribution of $L2_1$ and $A2$ phases in the sample, revealing that the $A2$ disorder is mainly located at the upper part of the films in vicinity to the film surface.\\

\section*{Acknowledgement}

We thank A. Alfonsov and A. Omar for fruitful discussions. 
Financial support is acknowledged from the Deutsche Forschungsgemeinschaft (DFG) through the Sonderforschungsbereich
SFB 1143 (project C02), and Grants No. WU595/3-3, WU595/14-1, and BA5656/1-1. Irradiation experiments were performed at the Ion Beam Center of the Helmholtz-Zentrum Dresden-Rossendorf  (HZDR). Furthermore, the use of HZDR Ion Beam Center TEM facilities and the funding of TEM Talos by the German Federal Ministry of Education of Research (BMBF), Grant No. 03SF0451 in the framework of HEMCP are acknowledged.\\


\end{document}